\documentclass[english,twocolumn,transaction,10pt]{IEEEtran}
\usepackage[T1]{fontenc}
\usepackage{float}
\usepackage{amsmath}
\usepackage{amsthm}
\usepackage{amssymb}
\usepackage{stackrel}
\usepackage{graphicx}
\usepackage{setspace}
\usepackage{url}
\usepackage{bm}
\usepackage{caption}
\usepackage{cases}
\usepackage{xcolor}
\usepackage{diagbox}
\usepackage{algorithm}
\usepackage{algorithmic}
\usepackage{setspace}
\usepackage{stfloats}
\allowdisplaybreaks[4]
\renewcommand{\algorithmicrequire}{ \textbf{Input:}} 
\renewcommand{\algorithmicensure}{ \textbf{Output:}} 
\usepackage{multirow}
\usepackage{makecell}
\usepackage{threeparttable}
\usepackage{setspace}

\makeatletter

\floatstyle{ruled}
\newfloat{algorithm}{tbp}{loa}
\providecommand{\algorithmname}{Algorithm}
\floatname{algorithm}{\protect\algorithmname}

\theoremstyle{plain}

\theoremstyle{definition}

\theoremstyle{plain}

\theoremstyle{plain}

\IEEEoverridecommandlockouts
\usepackage{ifpdf}
\usepackage{cite}
\usepackage{amsmath}
\usepackage{mathtools}
\usepackage{booktabs}
\usepackage{multirow}
\usepackage{setspace}
\usepackage{lettrine}
 \usepackage{array}
\@ifundefined{showcaptionsetup}{}{
	\PassOptionsToPackage{caption=false}{subfig}}
\usepackage{subfig}
\makeatother
\usepackage{babel}

\def\BibTeX{{\rm B\kern-.05em{\sc i\kern-.025em b}\kern-.08em
    T\kern-.1667em\lower.7ex\hbox{E}\kern-.125emX}}

\begin{document}

%

\title{ARIS-RSMA Enhanced ISAC System: Joint Rate Splitting and Beamforming Design
}

\author{Xin Jin, Tiejun Lv, \IEEEmembership{Senior Member,~IEEE}, Yashuai Cao, Jie Zeng, \IEEEmembership{Senior Member,~IEEE}, \\ and Mugen Peng, \IEEEmembership{Fellow,~IEEE} 


\thanks{Manuscript received 27 December 2025; accepted 5 February 2026. This paper was supported by the National Natural Science Foundation of China under No. 62271068. (\emph{corresponding author: Tiejun Lv}.)}

\thanks{X. Jin, T. Lv and M. Peng are with the School of Communication and Information Engineering, Beijing University of Posts and Telecommunications, Beijing 100876, China (e-mail: \{jxzoe, lvtiejun, pmg\}@bupt.edu.cn). 

Y. Cao is with the School of Intelligence Science and Technology, University of Science and Technology Beijing, Beijing 100083, China (e-mail: caoys@ustb.edu.cn). 

Jie Zeng is with the School of Cyberspace Science and Technology, Beijing Institute of Technology, Beijing 100081, China (e-mail: zengjie@bit.edu.cn).}

}

\maketitle

\begin{abstract}
This letter proposes an active reconfigurable intelligent surface (ARIS) assisted rate-splitting multiple access (RSMA) integrated sensing and communication (ISAC) system to overcome the fairness bottleneck in multi-target sensing under obstructed line-of-sight environments. Beamforming at the transceiver and ARIS, along with rate splitting, are optimized to maximize the minimum multi-target echo signal-to-interference-plus-noise ratio under multi-user rate and power constraints. The intricate non-convex problem is decoupled into three subproblems and solved iteratively by majorization-minimization (MM) and sequential rank-one constraint relaxation (SROCR) algorithms. Simulations show our scheme outperforms non-orthogonal multiple access, space-division multiple access, and passive RIS baselines, approaching sensing-only upper bounds.
\end{abstract}


\IEEEpeerreviewmaketitle

\section{Introduction}

\IEEEPARstart{I}{ntegrated} sensing and communication (ISAC) signifies a paradigm shift by integrating communication and sensing functions within a unified platform to enhance hardware and spectrum efficiency \cite{10839626}. However, practical ISAC deployments remain challenged by obstructed line-of-sight (LoS) coverage limitations, intricate interference management, and the fairness bottleneck in sensing multiple targets.

To address the coverage issue, active reconfigurable intelligent surface (ARIS) has been employed to amplify weak incident signals \cite{10319318}. Current studies on ARIS-ISAC typically optimize beamforming for sensing signal-to-interference-plus-noise ratio (SINR) maximization \cite{10054402} or energy efficiency optimization \cite{10685084}, primarily under single-target scenarios based on the space-division multiple access (SDMA) scheme. While ARIS enhances link budget, the interference nulling strategy inherent to SDMA results in sub-optimal spatial resource utilization. This inefficiency prevents the system from achieving balanced echo SINR across multiple targets, leading to a critical fairness bottleneck for sensing.

Rate-splitting multiple access (RSMA), which treats part of the interference as a decodable stream, provides a more flexible and robust interference management strategy \cite{10038476}. RSMA-based ISAC designs have been shown to improve the communication-sensing trade-off \cite{9531484}, and to enhance multi-user fairness compared to SDMA \cite{10486996}. Furthermore, RSMA has been combined with transmissive metasurfaces to enable secure ISAC networks \cite{10945425}. However, these studies typically assume favorable propagation conditions. In LoS-obstructed environments, advanced interference management alone cannot overcome the physical coverage deficit. Recently, ARIS-assisted RSMA-ISAC frameworks have been explored for security-oriented designs under eavesdropper spatial uncertainty \cite{11165755} and power transfer \cite{11122502}. While these studies demonstrate the mutual benefits of ARIS and RSMA, the critical problem of fairness-guaranteed multi-target sensing under blockage has yet to be explored.

To bridge this gap, we propose an integrated ARIS-RSMA ISAC framework to circumvent the resource competition bottleneck between multi-user communication and multi-target sensing. Specifically, we leverage RSMA to bypass the resource-intensive nulling requirements of SDMA, effectively harnessing spatial resources for fair multi-target sensing. However, this integration introduces challenges, including the intricate coupling between the ARIS amplification budget and the RSMA stream design, the cascaded non-convexity from the round-trip ARIS structure, and the dual impact of amplified noise on both communication reliability and echo quality.

In this letter, we jointly design the base station (BS) transmit and receive beamformers, ARIS reflection coefficients, and rate-splitting strategy to maximize the minimum multi-target echo SINR under multi-user quality of service (QoS) and power constraints. To resolve this highly coupled problem, a structured block coordinate descent (BCD)-based algorithm is developed. Numerical results validate the superiority of the proposed method over the non-orthogonal multiple access (NOMA), SDMA, and passive RIS (PRIS) baselines, with performance approaching the sensing-only upper bounds.

\section{System Model And Problem Statement}

\begin{figure}[t]
	\centering
	\includegraphics[scale=0.3]{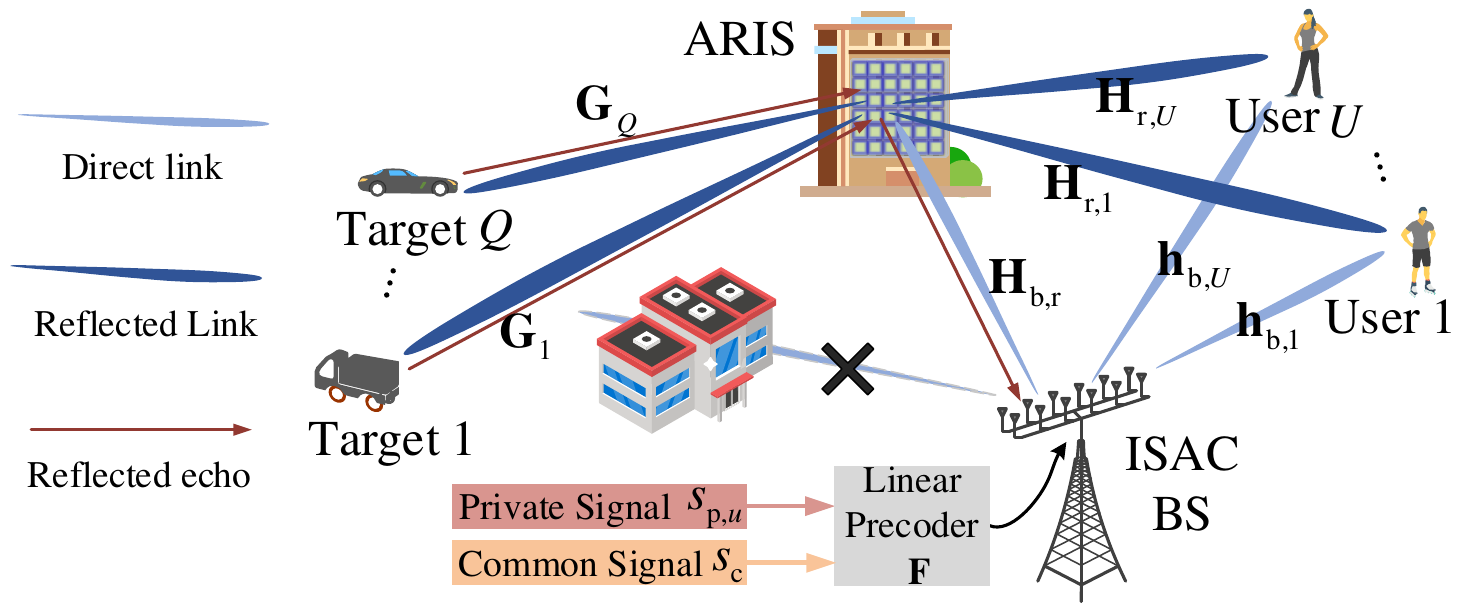}    \captionsetup{justification=raggedright,font={small}}
	\caption{Illustration of the ARIS-RSMA ISAC system.} 
	\label{fig:sysmodel}
\end{figure}

As shown in Fig. \ref{fig:sysmodel}, we consider an ARIS-RSMA enhanced ISAC system. The monostatic ISAC BS configured with $M$ transmit/receive antennas simultaneously performs $U$ single-antenna user downlink communication tasks, $\mathcal{U} = \{1,…,U\}$, and $Q$ target sensing tasks, $\mathcal{Q} = \{1,…,Q\}$. Given that target detection strongly relies on the LoS link between BS and targets, an $L$-element ARIS, $\mathcal{L} = \{1,…,L\}$, is deployed in the ISAC system with blocked targets to enhance sensing and communication performance. Define the reflection coefficient of the $l$-th ARIS element as ${\varphi _l} = {a _l}{e^{\iota {\theta _l}}}$, where ${a _l}$ is the amplification and ${\theta _l}$ is the phase shift. The ARIS beamforming vector is then given by ${\bm{\varphi }} = {[ {{\varphi _1},...,{\varphi _L}} ]^T}$. 

In the RSMA-assisted ISAC system, the transmit data streams comprise private streams for $U$ users, ${s_{{\rm{p}},u}}, u \in \mathcal{U}$, and a common stream $s_{\rm{c}}$. The message-splitting strategy for private and common streams in RSMA is similar to \cite{9531484}. Denoting the compound data streams ${\bf{s}} \!=\! {\left[ {{s_{{\rm{p}},1}},...,{s_{{\rm{p}},U}},{s_{\rm{c}}}} \right]^T} \! \in \! \mathbb{C}{^{U \!+\! 1}}$, with $\mathbb{E} (\! {{\bf{s}}{{\bf{s}}\!^H}} \!) \! =\! {\bf{I}}_{U\!+\!1}$, the transmit signal is given by
\begin{equation} \label{xsignal}
    {\bf{x}} = {\bf{Fs}} = \textstyle\sum_{u \in {\mathcal{U}}} {{{\bf{f}}_{{\rm{p}},u}}{s_{{\rm{p}},u}}}  + {{\bf{f}}_{\rm{c}}} {s_{\rm{c}}},
\end{equation}
where ${\bf{F}} \!=\! \left[ { {{\bf{f}}_{{\rm{p}},1}},...,{{\bf{f}}_{{\rm{p}},U}}, {{\bf{f}}_{\rm{c}}} } \right] \! \in \! \mathbb{C}{^{M \!\times\! \left( {U \!+\! 1} \right)}}$. Each column of ${\bf{F}}$ corresponds to the transmit beamforming vector of ${\bf{s}}$.

Let ${{\bf{H}}_{{\rm{b}},{\rm{r}}}} \in \mathbb{C}^{L \times M}$, ${{\bf{h}}_{{\rm{r}},u}} \in \mathbb{C}^{L }$, and ${{\bf{h}}_{{\rm{b}},u}} \in \mathbb{C}^{M }$ denote the channel between BS and ARIS, ARIS and user $u$, BS and user $u$, respectively. The received signal $y_u$ at user $u$ is given as
\begin{equation} \label{y_u}
    {y_u} = \left( {{\bf{h}}_{{\rm{b}},u}^H + {\bf{h}}_{{\rm{r}},u}^H{\bf{\Phi }}{{\bf{H}}_{{\rm{b}},{\rm{r}}}}} \right){\bf{x}} + {\bf{h}}_{{\rm{r}},u}^H{\bf{\Phi }}{{\bf{z}}_1} + {n_u},
\end{equation}
with ${\bf{\Phi }} = {\rm{diag}}( {\bm{\varphi }})$. ${{\bf{z}}_1} \sim {\cal CN}\left( {{\bf{0}},\sigma _{\rm{z}}^2{{\bf{I}}_L}} \right)$ and ${n_u} \sim {\cal CN}\left( {0,\sigma _u^2} \right)$ are the dynamic noise at ARIS and the additive white Gaussian noise (AWGN) at user $u$, respectively. For simplicity, we define ${\bf{h}}_u^H = {{\bf{h}}_{{\rm{b}},u}^H + {\bf{h}}_{{\rm{r}},u}^H{\bf{\Phi }}{{\bf{H}}_{{\rm{b}},{\rm{r}}}}}$ as the equivalent channel between the BS and user $u$.

Each user first extracts the common stream $s_{\rm{c}}$ from $y_u$ by treating all private streams as interference \cite{9531484}. After removing $s_{\rm{c}}$ via successive interference cancellation (SIC), user $u$ decodes $s_{{\rm{p}},u}$ with the remaining private streams treated as interference. Hence, the SINR of decoding $s_{\rm{c}}$ and $s_{{\rm{p}},u}$ at user $u$ are expressed as
\begin{equation} \label{gamma_c}
    {\gamma _{{\rm{c}},u}} \!=\! {{{{| {{\bf{h}}_u^H{{\bf{f}}_{\rm{c}}}} |}^2}}} / \left({{\textstyle\sum_{k \in {\mathcal U}} \!{{{| {{\bf{h}}_u^H{{\bf{f}}_{{\rm{p}},k}}} |}^2}  }  \!+\! \sigma _{\rm{z}}^2{{\| {{\bf{h}}_{{\rm{r}},u}^H{\bf{\Phi }}} \|}^2} \! +\! \sigma _u^2}} \right),
\end{equation}
\begin{equation}  \label{gamma_p}
    {\gamma _{{\rm{p}},u}} \!=\! {{{{| {{\bf{h}}_u^H {{\bf{f}}_{{\rm{p}},u}}} |}^2}}}/ \left( \!{{\textstyle\sum_{k \in {\tilde{\mathcal{U}}}}\! {{{| {{\bf{h}}_u^H{{\bf{f}}_{{\rm{p}},k}}} |}^2} } \!+\! \sigma_{\rm{z}}^2{{\| {{\bf{h}}_{{\rm{r}},u}^H \!{\bf{\Phi }}} \|}^2} \!+\! \sigma_u^2}} \!\right),
\end{equation}
with $\tilde{\mathcal{U}} \!\!=\!\! {\mathcal{U}} \backslash u$. The corresponding achievable rates are given by ${r_{{\rm{c}},u}} \!=\! {\log _2}( {1 \!+\! {\gamma _{{\rm{c}},u}}} )$ and ${r_{{\rm{p}},u}} \!=\! {\log _2}( {1 \!+\! {\gamma _{{\rm{p}},u}}} )$, respectively.

As the targets are blocked, ISAC BS senses $Q$ targets through ARIS-assisted equivalent LoS paths. Thus, the echo signal undergoing the BS-ARIS-target-ARIS-BS path is
\begin{equation} \label{y_r}
    {{\bf{y}}_{\rm{r}}} = {\bf{H}}_{{\rm{b}},{\rm{r}}}^H{{\bf{\Phi }}^H}( {\textstyle\sum_{q \in {\mathcal{Q}}} {{\bf{G}}_q} {\bf{\Phi }}\left( {{{\bf{H}}_{{\rm{b}},{\rm{r}}}}{\bf{x}} + {{\bf{z}}_1}} \right) + {{\bf{z}}_2}}) + {{\bf{n}}_{\rm{r}}},
\end{equation}
where ${{\bf{G}}_q} \!\!=\!\! {\beta _q}{\bf{a}}( \!{{\phi _q}} \!){{\bf{a}}^H}\!(\! {{\phi _q}} \!)$ is the target response matrix between ARIS and target $q$. ${\beta _q}$ is the channel gain coefficient containing path loss and radar cross section (RCS), and ${\bf{a}}\left( {{\phi _q}} \right) = {[1,{e^{\iota \pi \sin \phi_q }},...,{e^{\iota (L - 1)\pi \sin \phi_q }}]^T}$ is the array steering vector of ARIS concerning the target direction ${\phi _q}, q \in \mathcal{Q}$. ${{\bf{z}}_2} \sim {\cal CN}\left( {{\bf{0}},\sigma _{\rm{z}}^2{{\bf{I}}_L}} \right)$ and ${{\bf{n}}_{\rm{r}}} \sim {\cal C}{\cal N}( {{\bf{0}},\sigma _{\rm{r}}^2{{\bf{I}}_M}} )$ represent the dynamic noise at ARIS and the AWGN, respectively. 

Multi-target detection and parameter estimation based on RIS is achievable; hence, we assume that all channels are perfectly known \cite{10054402}. To ensure reliable detection in obstructed multi-target scenarios, we adopt the max-min echo SINR criterion. This criterion directly couples with detection probability, offers superior tractability for joint beamforming optimization compared to the Cram\'{e}r-Rao bound (CRB), and provides worst-case performance guarantees \cite{10685084}. A receive beamformer, ${\bf{w}}_q \in \mathbb{C}^{M}$, is employed to enhance the echo SINR ${\gamma _{{\rm{r}},q}}$ of target $q$, thus ${\gamma _{{\rm{r}},q}}$ can be written as
\begin{align} \label{gamma_q}
    {\gamma _{{\rm{r}},q}} &= \frac{{{{| {{\bf{w}}_q^H{{\bf{H}}_{{\rm{b}},q}}{\bf{Fs}}} |}^2}}}{{{{| {{\bf{w}}_q^H{{{\bf{\tilde H}}}_{{\rm{b}},q}}{\bf{Fs}}} |}^2} + {{| {{\bf{w}}_q^H\left( {{{\bf{H}}_{z_1}}{{\bf{z}}_1} + {{\bf{H}}_{z_2}}{{\bf{z}}_2} + {{\bf{n}}_{\rm{r}}}} \right)} |}^2}}} \nonumber \\ 
    &= \frac{{{\bf{w}}_q^H{{\bf{H}}_{{\rm{b}},q}}{\bf{F}}{{\bf{F}}^H}{\bf{H}}_{{\rm{b}},q}^H{{\bf{w}}_q}}}{{{\bf{w}}_q^H( {{{{\bf{\tilde H}}}_{{\rm{b}},q}}{\bf{F}}{{\bf{F}}^H}{\bf{\tilde H}}_{{\rm{b}},q}^H + {\bf{C}}} ){{\bf{w}}_q}}},  
\end{align}
where ${{\bf{H}}_{{\rm{b}},q}} = {\bf{H}}_{{\rm{b}},{\rm{r}}}^H{{\bf{\Phi }}^H}{{\bf{G}}_q}{\bf{\Phi }}{{\bf{H}}_{{\rm{b}},{\rm{r}}}}$, ${{{\bf{\tilde H}}}_{{\rm{b}},q}} = {\bf{H}}_{{\rm{b}},{\rm{r}}}^H{{\bf{\Phi }}^H} {{\bf{G}}_{\tilde{q}}} {\bf{\Phi }}{{\bf{H}}_{{\rm{b}},{\rm{r}}}}$ with ${{\bf{G}}_{\tilde{q}}} = \sum_{j \in Q,j \ne q} {{{\bf{G}}_j}}$, ${\bf{G}} = \sum_{q \in \mathcal{Q}} {{\bf{G}}_q}$, ${{\bf{H}}_{z_1}} = {\bf{H}}_{{\rm{b}},{\rm{r}}}^H{{\bf{\Phi }}^H}{\bf{G\Phi }}$, ${{\bf{H}}_{z_2}} = {\bf{H}}_{{\rm{b}},{\rm{r}}}^H{{\bf{\Phi }}^H}$, and ${\bf{C}} = \sigma _{\rm{z}}^2{{\bf{H}}_{z_1}}{\bf{H}}_{z_1}^H + \sigma _{\rm{z}}^2{{\bf{H}}_{z_2}}{\bf{H}}_{z_2}^H + \sigma _{\rm{r}}^2{{\bf{I}}_M}$.

Our goal is to maximize the worst-case echo SINR over all targets by jointly designing the receive beamformers $\{{\bf w}_q\}$, the BS transmit beamformer ${\bf F}$, the ARIS reflection matrix ${\bf \Phi}$, and the RSMA common rate allocation ${\bf{c}} = [c_1,...,c_U]^T \in \mathbb{R}_+^U$,  under user's QoS and system power constraints, i.e.,
\begin{subequations}
\begin{align}
    \mathrm{P}_0 :     
    \mathop {\max }\limits_{\left\{ {{{\bf{w}}_q}} \right\}, {\bf{F}}, {\bf{\Phi }}, {\bf{c}} } 
    &\mathop {\min }\limits_{q \in Q}  {\gamma _{{\rm{r}},q}}  
    \label{eq:problem0A} \\
    \mathrm{s.t.} \ & {c_u} + {r_{{\rm{p}},u}} \ge R_u^{\min },\forall u \in {\cal U}, \label{eq:problem0B}\\
    & \textstyle\sum_{k \in {\cal U}} {{c_k}}  \le {r_{{\rm{c}},u}},  \forall u \in {\cal U}, \label{eq:problem0C} \\
    & \|{\bf{F}} \|_F^2 \le P_{{\rm{BS}}}^{{\rm{max}}}   \label{eq:problem0D} \\  
    & {P_{{\rm{RIS}}}} \le P_{{\rm{RIS}}}^{{\rm{max}}},   \label{eq:problem0E} \\
    &{a _l} \le {a_{\rm{max} }},\forall l \in \mathcal{L}.  \label{eq:problem0F}
\end{align}
\end{subequations}
The vector $\mathbf{c}$ is jointly optimized with the beamformers in problem $\mathrm{P}_0$ under the per-user QoS constraint \eqref{eq:problem0B} and the common-stream decodability constraint \eqref{eq:problem0C}. This ensures user fairness at a prescribed threshold level, while the remaining spatial degrees of freedom are steered to maximize the minimum echo SINR, thereby enhancing sensing fairness. $P_{{\rm{BS}}}^{{\rm{max}}}$ and $P_{{\rm{RIS}}}^{{\rm{max}}}$ specify the maximum power budget for ISAC BS and ARIS, respectively. ${a_{\rm{max} }}$ is the maximum amplification factor for each ARIS element. Particularly, ${P_{{\rm{RIS}}}}$ is the reflecting power consumption at ARIS, given by
\begin{equation}  \label{PRIS}
\begin{aligned}
 {P_{{\rm{RIS}}}} \!\!=\!\! {\left\| \!{{\bf{\Phi }}\!{{\bf{H}}_{{\rm{b}},{\rm{r}}}}\!{\bf{F}}} \! \right\|}\!_F^2 \!\!+\!\! {\left\| \!{{{\bf{\Phi }}\!^H}\!{\bf{G\Phi }}\!{{\bf{H}}_{{\rm{b}},{\rm{r}}}}\!{\bf{F}}}\! \right\|}\!_F^2 \!\!+\!\! \sigma _{\rm{z}}^2 \! {\left\| \!{{{\bf{\Phi }}\!^H}\!{\bf{G\Phi }}}\! \right\|}\!_F^2 \!\!+\!\! 2\sigma _{\rm{z}}^2 \! {\left\| \!{{{\bf{\Phi }}\!^H}}\! \right\|}\!_F^2.
\end{aligned}
\end{equation}

Solving problem $\mathrm{P}_0$ presents critical challenges: i) The fractional forms of the echo SINR in \eqref{eq:problem0A} and the user rate in \eqref{eq:problem0B} and \eqref{eq:problem0C}, introduce non-convexity; ii) the quartic objective \eqref{eq:problem0A} and constraint \eqref{eq:problem0E} further increase complexity;\! and iii) the intricate coupling between the transmit beamforming ${\bf{F}}$ and the reflection coefficient ${\bf{\Phi }}$ hinders the quest for a global optimal solution of problem $\mathrm{P}_0$. This motivates the development of a BCD-based iterative algorithm in Section III.

\section{Algorithm Design}

\subsection{Receive Beamforming Design}
Considering that the receive beamforming ${\bf{w}}_q$ is only related to the echo SINR in the objective function \eqref{eq:problem0A}, we design ${\bf{w}}_q$ by maximizing the echo SINR for target $q$, written as
\begin{equation} 
    \mathrm{P}_1 :
    {\bf{w}}_q^{{\rm{opt}}} = \arg \mathop {\max }\limits_{{{\bf{w}}_q}} \frac{{{\bf{w}}_q^H{{\bf{A}}_1}{{\bf{w}}_q}}}{{{\bf{w}}_q^H{{\bf{A}}_2}{{\bf{w}}_q}}}, \forall q \in \mathcal{Q}, \label{eq:problem1A}
\end{equation}
where ${{\bf{A}}_1} = {{\bf{H}}_{{\rm{b}},q}}{\bf{F}}{{\bf{F}}^H}{\bf{H}}_{{\rm{b}},q}^H$, and ${{\bf{A}}_2} = {{{\bf{\tilde H}}}_{{\rm{b}},q}}{\bf{F}}{{\bf{F}}^H}{\bf{\tilde H}}_{{\rm{b}},q}^H + {\bf{C}}$. This problem is a standard generalized Rayleigh quotient optimization problem, thereby the optimal solution ${\bf{w}}_q^{{\rm{opt}}}$ is the eigenvector affiliated with the maximum eigenvalue of ${\bf{A}}_2^{ - 1}{{\bf{A}}_1}$. At the $t$-th iteration, the optimal minimum multi-target  SINR after solving the problem $\mathrm{P}_1$ is given by
\begin{equation} \label{Gamma_P1}
    \{ {{\varGamma ^{{\rm{opt}}}}} \}_{{{\rm{P}}_1}}^t = \mathop {\min }\limits_q \frac{{{{( {{\bf{w}}_q^{{\rm{opt}}}})\!}^H}{{\bf{A}}_1}{\bf{w}}_q^{{\rm{opt}}}}}{{{{( {{\bf{w}}_q^{{\rm{opt}}}} )\!}^H}{{\bf{A}}_2}{\bf{w}}_q^{{\rm{opt}}}}}.
\end{equation}

\subsection{Joint Transmit Beamforming and Rate-Splitting Design}
Given ${\bf{w}}_q$ and ${\bf{\Phi}}$, we reformulate the subproblem of jointly optimizing the transmit beamforming ${\bf{F}}$ and rate splitting $\bf{c}$ as
\begin{subequations}
\begin{align}
    \mathrm{P}_2 :     
    \mathop {\max }\limits_{ {\bf{c}}, \{{\bf{f}}_{{\rm{p}},u}\}, {\bf{f}}_{\rm{c}} }
    &\mathop {\min }\limits_{q \in Q}  \frac{{\textstyle\sum_{i \in {\cal I}} {{\rm{tr}}( {{{\bf{B}}_1}{{\bf{f}}_i}{\bf{f}}_i^H} )} }}{{\textstyle\sum_{i \in {\cal I}} {{\rm{tr}}( {{{\bf{B}}_2}{{\bf{f}}_i}{\bf{f}}_i^H} )}  + {\varepsilon _1}}}   
    \label{eq:problem2A} \\
    \mathrm{s.t.} \  & \textstyle\sum_{i \in {\cal I}} {{\rm{tr}}\left( {{{\bf{f}}_i}{\bf{f}}_i^H} \right)}  \le P_{{\rm{BS}}}^{{\rm{max}}}, \label{eq:problem2B} \\
    & \textstyle\sum_{i \in {\cal I}} {{\rm{tr}}\left( {{\bf{\Sigma }}{{\bf{f}}_i}{\bf{f}}_i^H} \right)} + {\varepsilon _2}  \le P_{{\rm{RIS}}}^{{\rm{max}}}, \label{eq:problem2C} \\   &\eqref{eq:problem0B},\eqref{eq:problem0C}, \label{eq:problem2D} 
\end{align}
\end{subequations}
where $\mathcal{I} \!\!=\!\! \{1,...,U,{\rm{c}}\}$, and ${\bf{f}}_i$ is the $i$-th column of $\bf{F}$. ${{\bf{B}}_1} \! = \!{\bf{H}}_{{\rm{b}},q}^H{{\bf{w}}_q}{\bf{w}}_q^H{{\bf{H}}_{{\rm{b}},q}}$, ${{\bf{B}}_2} \!=\! {\bf{\tilde H}}_{{\rm{b}},q}^H{{\bf{w}}_q}{\bf{w}}_q^H{{{\bf{\tilde H}}}_{{\rm{b}},q}}$, ${\varepsilon _1} \!=\! {\bf{w}}_q^H{\bf{C}}{{\bf{w}}_q}$, ${\bf{\Sigma }} \!=\! {{\bf{H}}_{z_1}}{\bf{H}}_{z_1}^H \!+\! {{\bf{H}}_{z_2}}{\bf{H}}_{z_2}^H$, and ${\varepsilon _2} \!=\! \sigma _{\rm{z}}^2\left\| {{{\bf{\Phi }}^H}{\bf{G\Phi }}} \right\|_F^2 \!+\! 2\sigma _{\rm{z}}^2\left\| {{{\bf{\Phi }}^H}} \right\|_F^2$. 

Problem $\mathrm{P}_2$ is a non-convex fractional quadratically constrained quadratic programming (QCQP). To handle the quadratic beamforming terms, we lift the vector variables by defining ${{\bf{F}}_i} = {{\bf{f}}_i}{\bf{f}}_i^H$, $\forall i \in {\cal I}$, and ${{\bf{H}}_u} = {{\bf{h}}_u}{\bf{h}}_u^H$, $\forall u \in {\cal U}$. The rate constraints \eqref{eq:problem0B} and \eqref{eq:problem0C} are then written as
\begin{equation} \label{eq:rate_u1}
    {c_u} + {\log _2}( {1 + \frac{{{\rm{tr}}( {{{\bf{H}}_u}{{\bf{F}}_u}} )}} {{\textstyle\sum_{ k \in {\tilde{\mathcal{U}}}} {{\rm{tr}}( {{{\bf{H}}_u}{{\bf{F}}_k}} )}  + {\varepsilon _3}}} } )  \ge  R_u^{\rm{min} },
\end{equation}
\begin{equation} \label{eq:rate_c1}
    \textstyle\sum_{k \in {\cal U}} {{c_k}}  \le {\log _2}( {1 + \frac{{{\rm{tr}}( {{{\bf{H}}_u}{{\bf{F}}_{\mathrm{c}}}} )}}{{\sum_{ k \in {\mathcal{U}}} {{\rm{tr}}( {{{\bf{H}}_u}{{\bf{F}}_k}} )}  + {\varepsilon _3}}} } ).
\end{equation}
with $\varepsilon _3 \!=\! \sigma_{\rm{z}}^2{{\| {{\bf{h}}_{{\rm{r}},u}^H \!{\bf{\Phi }}} \|}^2} \!+\! \sigma_u^2$. Subsequently, we introduce $\{ \varGamma \}_{{{\rm{P}}_x}}^t$ as the minimum multi-target SINR when solving problem $\mathrm{P}_x$ in the $t$-th iteration, and reformulate the problem $\mathrm{P}_2$ as 
\begin{subequations}
\begin{align}
    \mathop {\max }\limits_{ {\bf{c}}, \{{{\bf{F}}_i}\succeq 0\}, \{\! \varGamma \!\}_{{{\rm{P}}_{2}}}^t} 
    &\mathop \{ \varGamma \}_{{{\rm{P}}_{2}}}^t       \label{eq:problem2-1A} \\
    \mathrm{s.t.} \ &\textstyle\sum_{i \in {\cal I}} \!{{\rm{tr}} \! \left(  {( {{{\bf{B}}_1} \!-\! \{ {\varGamma} \}_{{{\rm{P}}_{2}}}^t{{\bf{B}}_2}} ){{\bf{F}}_i}} \right)\!} \ge \{ \varGamma\}_{{{\rm{P}}_{2}}}^t{\varepsilon _1},\nonumber \\ 
    & \ \forall q \!\in\! {\cal Q}, \label{eq:problem2-1B}\\ 
    & \textstyle\sum_{i \in {\cal I}} {{\rm{tr}}\left( {{{\bf{F}}_i}} \right)}  \le P_{{\rm{BS}}}^{{\rm{max}}}, \label{eq:problem2-1C}\\
    & \textstyle\sum_{i \in {\cal I}} {{\rm{tr}}\left( {{\bf{\Sigma }}{{\bf{F}}_i}} \right)} \le P_{{\rm{RIS}}}^{{\rm{max}}} - {\varepsilon _2}, \label{eq:problem2-1D} \\    
    & {\rm{rank}}({{\bf{F}}_i}) = 1, \forall i \in \mathcal{I},  \label{eq:problem2-1E} \\
    & \eqref{eq:rate_u1},\eqref{eq:rate_c1}. \label{eq:problem2-1F} 
\end{align}
\end{subequations}

Due to non-convex constraints \eqref{eq:problem2-1B}, \eqref{eq:problem2-1E}, and \eqref{eq:problem2-1F}, this problem remains unsolvable directly. Inspired by the low-complexity SINR approximation algorithm in \cite{9163260}, we resort to approximate $\{ \varGamma \}_{{{\rm{P}}_{2}}}^t$ on the left-hand side of constraint \eqref{eq:problem2-1B} based on $\{ {{\varGamma ^{{\rm{opt}}}}} \}_{{{\rm{P}}_1}}^t$ of problem $\mathrm{P}_{1}$, i.e., 
\begin{equation} \label{eq:Gamma1}
\textstyle\sum_{i \in {\cal I}} \!{{\rm{tr}} \! \left(  {( {{{\bf{B}}_1} \!-\! \{ {{\varGamma ^{{\rm{opt}}}}} \}_{{{\rm{P}}_1}}^t {{\bf{B}}_2}} ){{\bf{F}}_i}} \right)\!} \ge \{ \varGamma\}_{{{\rm{P}}_{2}}}^t{\varepsilon _1}, \forall q \in \mathcal{Q}.
\end{equation}

By introducing a series of non-negative real variables, $\chi = \{\rho _{{\rm{p}},u}$, $\rho _{{\rm{c}},u}$, $\xi _{{\rm{p}},u}$, $\xi _{{\rm{c}},u}\}$, $\forall u \!\in \!\mathcal{U}$, and utilizing first-order Taylor approximation techniques, the rate constraints in \eqref{eq:problem2-1F} are relaxed and approximated by
\begin{equation}  \label{eq:rate_u2}
    {c_u} + ( {{\rho _{{\rm{p,}}u}} - {\xi _{{\rm{p}},u}}} ) / {\ln 2}  \ge R_u^{\mathrm{min}},
\end{equation}
\begin{equation}  \label{eq:rho_p1}  
    \textstyle\sum_{k \in {\mathcal U}} {{\rm{tr}}\left( {{{\bf{H}}_u}{{\bf{F}}_k}} \right)}  + {\varepsilon _3} \ge {e^{{\rho _{{\rm{p}},u}}}},
\end{equation}
\begin{equation}   \label{eq:xi_p2}
    \textstyle\sum_{k \in \tilde{\mathcal U}} \!{{\rm{tr}}(\! {{{\bf{H}}_u}{{\bf{F}}_k}} \!)}  + {\varepsilon _3} \le {e^{\xi _{{\rm{p}},u}^{(\!t - 1\!)}}}( {{\xi _{{\rm{p}},u}} \!-\! \xi _{{\rm{p}},u}^{(\!t - 1\!)} \!+\! 1} ),
\end{equation}
\begin{equation} \label{eq:rate_c2}
    {\rho _{{\rm{c}},u}} - {\xi _{{\rm{c}},u}} \ge \ln 2 \textstyle\sum_{k \in {\cal U}} {{c_k}},
\end{equation}
\begin{equation}    \label{eq:rho_c1}
    \textstyle\sum_{i \in \mathcal{I}} {{\rm{tr}}\left( {{{\bf{H}}_u}{{\bf{F}}_i}} \right)}  + {\varepsilon _3} \ge {e^{{\rho _{{\rm{c}},u}}}},
\end{equation}
\begin{equation}   \label{eq:xi_c2}
    \textstyle\sum_{k \in {\mathcal U}} {{\rm{tr}}( {{{\bf{H}}_u}{{\bf{F}}_k}})}  + {\varepsilon _3} \le {e^{\xi _{{\rm{c}},u}^{(\!t - 1\!)}}}( {{\xi _{{\rm{c}},u}} \!-\! \xi _{{\rm{c}},u}^{(\!t - 1\!)} \!+\! 1} ).
\end{equation} 

For the rank-one constraint in \eqref{eq:problem2-1E}, rather than completely removing it, we employ sequential rank-one constraint relaxation (SROCR) to progressively tighten the semidefinite relaxation (SDR) solution \cite{8081370}. Using a flexible parameter $\varpi_i$, $\varpi_i \in [0, 1]$, \eqref{eq:problem2-1E} can be substituted with
\begin{equation} \label{eq:rank1}
    {\bf{u}}_{\max }^H( {{\bf{F}}_i^{( {\!t \!-\! 1\!} )}} ){{\bf{F}}_i}{{\bf{u}}_{\max }}( {{\bf{F}}_i^{( {\!t \!-\! 1\!})}} ) \ge \varpi _i^{( {\!t \!-\! 1\!} )}{\rm{tr}}( {{{\bf{F}}_i}} ),\forall i \in {\cal I},
\end{equation}
where ${{\bf{u}}_{\max }}(\cdot)$ is the largest eigenvector. We update
$\varpi_i^{(t)} = \min(1, \lambda_{\max}({\bf{F}}_i^{(t)})/{\rm{tr}}({\bf{F}}_i^{(t)}) \!+\! \delta_i^{(t)})$ with an adaptive step size $\delta_i^{(t)}$. Due to space limitations, specific steps and convergence of the SROCR algorithm can be found in \cite{8081370}.

To this end, the problem $\mathrm{P}_2$ is transformed into
\begin{subequations}
\begin{align}
    \mathop {\max }\limits_{ {\bf{c}}, \{{{\bf{F}}_i} \succeq\! 0\}, \{\! \varGamma \!\}_{{{\rm{P}}_{2}}}^t\!, \chi} 
    &\mathop \{ \varGamma \}_{{{\rm{P}}_{2}}}^t       \label{eq:problem2-2A} \\
    \mathrm{s.t.}  &\eqref{eq:problem2-1C},\! \eqref{eq:problem2-1D},\! \eqref{eq:Gamma1} \!-\! \eqref{eq:rank1},
    \label{eq:problem2-2B} 
\end{align}
\end{subequations}
which can be solved by the off-the-shelf toolboxes, e.g., CVX.

\subsection{ARIS Reflection Beamforming Design}
With the obtained ${\bf{w}}_q$, ${\bf{F}}$, $\bf{c}$, and the auxiliary variable $\{ \varGamma \}_{{{\rm{P}}_3}}^t$, the ARIS reflect beamforming subproblem is
\begin{subequations}
\begin{align}
    \mathrm{P}_{3} :     
    \mathop {\max }\limits_{ {\bf{\Phi}}, \{\! \varGamma \!\}_{{{\rm{P}}_{3}}}^t} 
    &\mathop \{ \varGamma \}_{{{\rm{P}}_{3}}}^t       \label{eq:problem3A} \\
    \mathrm{s.t.} \ & {\gamma _{{\rm{r}},q}} \ge \{ \varGamma \}_{{\rm{P}}_3}^t, \forall q \in {\cal Q}, \label{eq:problem3B}  \\    
    & \eqref{eq:problem0B},\eqref{eq:problem0C}, \eqref{eq:problem0E},\eqref{eq:problem0F}. \label{eq:problem3C}
\end{align}
\end{subequations}

The core difficulty in solving problem $\mathrm{P}_3$ is that the echo SINR and the ARIS power constraints contain quartic terms in ${\bm{\varphi}}$. This stems from the round-trip BS-ARIS-target-ARIS-BS echo, where ${\bf{\Phi}}\!=\!\rm{diag}({\bm{\varphi}})$ is involved twice. 

To make the problem tractable, we define ${\tilde{\bm{\varphi}}} \!=\! {\rm{vec}}( {{\bm{\varphi }}{{\bm{\varphi }}^H}})$ to reorganize element-pair interactions and reduce the dominant quartic terms to quadratic forms in ${\tilde{\bm{\varphi}}}$. Let $\bar{\bf{F}} = { {{{\bf{H}}_{{\rm{b}},{\rm{r}}}}{\bf{F}}{{\bf{F}}^H}\!{\bf{H}}_{{\rm{b}},{\rm{r}}}^H} }$, $\bar{\bf{W}} ={{\bf{H}}_{{\rm{b}},{\rm{r}}}}{{\bf{w}}_q}{\bf{w}}_q^H\!{\bf{H}}_{{\rm{b}},{\rm{r}}}^H$. Then, the echo SINR in \eqref{gamma_q} can be rewritten as
\begin{equation} \label{gamma_q2}
    {\gamma _{{\rm{r}},q}} \!=\! \frac{{\tilde{\bm{\varphi}}\!^H}\!{{\bf{M}}_1}\tilde{\bm{\varphi}}}{{{\tilde{\bm{\varphi}}}\!^H}\!({{\bf{M}}_2} \!+\! {\sigma _{\rm{z}}^2}\!{{\bf{M}}_3})\tilde{\bm{\varphi}} \!+\! \sigma _{\rm{z}}^2{{\bm{\varphi }}\!^H}\!{{\bf{M}}_4}{\bm{\varphi }} \!+\! \sigma _{\rm{r}}^2{{\bf{w}}_q^H}\!{{\bf{w}}_q}}, 
\end{equation}
where ${{\bf{M}}_1} \!=\! {{\bar{\bf{G}}}_q^H}\!( {{\bar{\bf{F}}^T}\! \! \otimes \! \bar{\bf{W}} } ) {{\bar{\bf{G}}}_q}$, ${{\bf{M}}_2} \!=\! {{\bar{\bf{G}}}_{\tilde{q}}^H}\!( {{\bar{\bf{F}}^T}\! \! \otimes \! \bar{\bf{W}} } ) {{\bar{\bf{G}}}_{\tilde{q}}}$, ${{\bf{M}}_3} \!=\! {{\bar{\bf{G}}}\!^H}\!( {{{\bf{I}}_L} \! \otimes \! \bar{\bf{W}} } ) {{\bar{\bf{G}}}}$, ${{\bf{M}}_4} \!=\! {{{\bf{I}}_L} \! \odot \! \bar{\bf{W}} }$, with 
${{\bar{\bf{G}}}_q} \!\!=\!\! {{\rm diag} (\! {{\rm{vec}} ( {{{\bf{G}}_q}})} \!)}$, ${{\bar{\bf{G}}}_{\tilde{q}}} \!\!=\!\! {{\rm diag} (\! {{\rm{vec}} ( {{{\bf{G}}_{\tilde{q}}}})} \!)}$, ${\bar{\bf{G}}} \!\!=\!\! {{\rm diag} (\! {{\rm{vec}} ( {\bf{G}})} \!)}$. 

Similar to the treatment of \eqref{eq:problem2-1B}, we approximate constraint \eqref{eq:problem3B} with $\{ \varGamma^{\rm{opt}}  \}_{{{\rm{P}}_2}}^t$, i.e., 
\begin{equation} \label{eq:Gamma2}
    {{\tilde{\bm{\varphi}}}^H}{\bf{M}}{\tilde{\bm{\varphi}}} \!+\! \{ \varGamma^{\rm{opt}}  \}_{{{\rm{P}}_2}}^t \sigma _{\rm{z}}^2{{\bm{\varphi }}^H}{{\bf{M}}_4}{\bm{\varphi }} \le \{ \varGamma  \}_{{{\rm{P}}_3}}^t{\tau _q}, \forall q \in {\cal Q},
\end{equation}
where ${\bf{M}} = \{ \varGamma^{\rm{opt}}  \}_{{{\rm{P}}_2}}^t \! ( {{{\bf{M}}_2} \!+\! \sigma _{\rm{z}}^2{{\bf{M}}_3}} )\! \!-\! {{\bf{M}}_1}$, ${\tau _q} = - \sigma _{\rm{r}}^2{{\bf{w}}_q^H}{{\bf{w}}_q}$. Recall \eqref{PRIS}, constraint \eqref{eq:problem0E} can be reorganized as
\begin{equation} \label{eq:PRIS}
    {{\tilde{\bm{\varphi}}}^H}{{\bf{D}}_1}{\tilde{\bm{\varphi}}} + {{\bm{\varphi }}^H}{{\bf{D}}_2}{\bm{\varphi }} \le P_{\rm{RIS}}^{\rm{max}},
\end{equation}
where ${{\bf{D}}_1} \!=\! {{\bar{\bf{G}}}\!^H}\!( {{\bar{\bf{F}}^T}\! \! \otimes \! {{\bf{I}}_L} } ) {\bar{\bf{G}}} \!+\! {\sigma _{\rm{z}}^2}{{\bar{\bf{G}}}\!^H}{\bar{\bf{G}}}$, ${{\bf{D}}_2} = ( {{{\bf{I}}_L} \! \odot \! {\bar{\bf{F}}} } ) \!+\! 2\sigma _{\rm{z}}^2{{\bf{I}}_L}$.

Based on \textit{Lemma 1} in \cite{10685084}, we approximate the higher-order terms in \eqref{eq:Gamma2} by majorization-minimization (MM), i.e., 
\begin{equation}
    {{\tilde{\bm{\varphi}}}\!^H}\!{\bf{M}}{\tilde{\bm{\varphi}}} \! \le \! {\lambda _{\rm{M}}}{{{\tilde{\bm{\varphi}}}}\!^H}\!{\tilde{\bm{\varphi}}} \!+\! 2\Re \!\{ {{{\tilde{\bm{\varphi}}}^H}\!(\!{\bf{M}} \! - \! {\lambda _{\rm{M}}}{\bf{I}}\!){{\tilde{\bm{\varphi}}}_s}} \}\! \! + \!{{\tilde{\bm{\varphi}}}_s^H}\!(\!{\lambda _{\rm{M}}}{\bf{I}} \!-\! {\bf{M}}\!){{\tilde{\bm{\varphi}}}_s},
\end{equation}
where ${\lambda _{\rm{M}}}$ is the maximum eigenvalue of $\mathbf{M}$, and ${{\tilde{\bm{\varphi}}}_s}$ is calculated by obtained ${\tilde{\bm{\varphi}}}$ in the $(t\!-\!1)$-th iteration. Given that ${a _l} \le a_\mathrm{max}$, we have ${{{\tilde{\bm{\varphi}}}}\!^H}\!{\tilde{\bm{\varphi}}} \le {L^2}a_{\max }^4$. Let ${\rm{vec}}( {{{\bf{P}}_1}} ) \!=\! {{\bf{p}}_1} \!=\! ({\bf{M}} \!-\! {\lambda _{\rm{M}}}{\bf{I}}){{\tilde{\bm{\varphi}}}_s}$, we further obtain
$2\Re \{ {{{\tilde{\bm{\varphi}}}\!^H}{{\bf{p}}_1}} \}  \!=\! 2\Re \{ {{{\bm{\varphi }}\!^H}{{\bf{P}}_1^H}{\bm{\varphi }}} \} \!=\! {{\bm{\varphi }}\!^H}( {{{\bf{P}}_1} \!+\! {{\bf{P}}_1^H}} ){\bm{\varphi }}$. Therefore, \eqref{eq:Gamma2} is converted into
\begin{equation} \label{eq:Gamma3}
    {{\bm{\varphi }}\!^H}{\bar{\bf{Q}}}_1{\bm{\varphi }} \!+\! {\eta _1} \! \le \! \{ \varGamma  \}_{{{\rm{P}}_3}}^t{\tau _q}, \forall q \! \in \!{\cal Q},
\end{equation}
with ${\bar{\bf{Q}}}_1 \!=\! {{{\bf{P}}_1} \!+\! {{\bf{P}}_1^H}} \!+\! \{\! \varGamma^{\rm{opt}}  \!\}_{{{\rm{P}}_2}}^t {\sigma _{\rm{z}}^2}{{\bf{M}}_4}$, ${\eta _1}\! =\! {\lambda _{\rm{M}}}{L^2}a_{\max }^4 \!+\! {{\tilde{\bm{\varphi}}}_s^H}(\!{\lambda _{\rm{M}}}{\bf{I}} \!-\! {\bf{M}}\!){{\tilde{\bm{\varphi}}}_s}$. Similarly, by defining ${\rm{vec}}( {{{\bf{P}}_2}} ) \!=\! {{\bf{p}}_2} \!=\! ({\bf{D}}_1 \!-\! {\lambda _{\rm{D}_1}}{\bf{I}}){{\tilde{\bm{\varphi}}}_s}$ with ${\lambda _{\rm{D}_1}}$ being the maximum eigenvalue of ${\bf{D}}_1$, \eqref{eq:PRIS} is replaced by
\begin{equation} \label{eq:PRIS2}
    {{\bm{\varphi }}\!^H}{{\bar{\bf{Q}}}_2}{\bm{\varphi }} \!+\! {\eta _2} \! \le \! P_{\rm{RIS}}^{\rm{max}},
\end{equation}
where ${\bar{\bf{Q}}}_2 \!\!=\!\! {{{\bf{P}}_2} \!+\! {{\bf{P}}_2^H}} \!+\! {{\bf{D}}_2}$, ${\eta _2}\! =\! {\lambda _{\rm{D}_1}}\!{L^2}a_{\max }^4 \!+\! {{\tilde{\bm{\varphi}}}_s^H}\!(\!{\lambda _{\rm{D}_1}}\!{\bf{I}} \!-\! {\bf{D}}_1\!){{\tilde{\bm{\varphi}}}_s}$.

Next, we address the user rate constraints. Since ${\bf{\Phi }}$ is diagonal, we have ${\bf{\Phi }}{{\bf{H}}_{{\rm{b}},{\rm{r}}}}{{\bf{f}}_{{\rm{p}},u}} \!= \!{\rm diag}( {{{\bf{H}}_{{\rm{b}},{\rm{r}}}}{{\bf{f}}_{{\rm{p}},u}}} ){\bm{\varphi }} \!=\! {{\tilde{\bf{H}}}_u}{\bm{\varphi }}$, and ${\bf{\Phi }}{{\bf{H}}_{{\rm{b}},{\rm{r}}}}{{\bf{f}}_{{\rm{c}}}} = {\rm diag}( {{{\bf{H}}_{{\rm{b}},{\rm{r}}}}{{\bf{f}}_{\rm{c}}}} ){\bm{\varphi }} = {{\tilde{\bf{H}}}_{\rm{c}}}{\bm{\varphi }}$. Therefore, the SINR can be reformulated as a quadratic function of $\bm{\varphi}$, i.e.,
\begin{equation}
    {\gamma _{x,u}} \!\!=\!\! \frac{{{{\bm{\varphi }}\!^H}{{\bf{E}}_{x_1}}{\bm{\varphi }} \!\!+\!\! 2\Re ( {{{\bf{E}}_{x_2}}{\bm{\varphi }}} ) \!\!+\!\! {\tau _{x_1}}}}{{{{\bm{\varphi }}\!^H}({{\bf{E}}_{x_3}}\!\!+\!\! {\sigma _{\rm{z}}^2}{{\bf{E}}_{x_4}} ){\bm{\varphi }} \!\!+\!\! 2\Re ( {{{\bf{E}}_{x_5}}{\bm{\varphi }}} ) \!\!+\!\! {\tau _{x_2}}}}, x\!\in\!\{{\rm{p}}/{\rm{c}}\},
\end{equation}
where ${{\bf{E}}_{{\rm{p}}_1}} \!=\!{{\tilde{\bf{H}}}_u^H}{{\bf{h}}_{{\rm{r}},u}}{{\bf{h}}_{{\rm{r}},u}^H} {{\tilde{\bf{H}}}_u}$, ${{\bf{E}}_{{\rm{p}}_2}} \!=\!{\bf{f}}_{{\rm{p}},u}^H{{\bf{h}}_{{\rm{b}},u}}{{\bf{h}}_{{\rm{r}},u}^H} {{\tilde{\bf{H}}}_u}$,  ${{\bf{E}}_{{\rm{p}}_3}} \!=\! \textstyle\sum_{k \in \tilde{\mathcal U} } {{\tilde{\bf{H}}}_k^H}{{\bf{h}}_{{\rm{r}},u}}{{\bf{h}}_{{\rm{r}},u}^H} {{\tilde{\bf{H}}}_k}$, ${{\bf{E}}_{{\rm{p}}_4}} \!=\! {\rm{diag}} {(\! {{{\bf{h}}_{{\rm{r}},u}}} \!)\!^H}{\rm{diag}} (\! {{{\bf{h}}_{{\rm{r}},u}}} \!)$, ${{\bf{E}}_{{\rm{p}}_5}} \!=\! \textstyle\sum_{k \in \tilde{\mathcal U}} {\bf{f}}_{{\rm{p}},k}^H{{\bf{h}}_{{\rm{b}},u}}{{\bf{h}}_{{\rm{r}},u}^H} {{\tilde{\bf{H}}}_k}$, ${\tau _{{\rm{p}}_1}} = {|{{\bf{h}}_{{\rm{b}},u}^H{{\bf{f}}_{{\rm{p}},u}}} |^2}$, ${\tau _{{\rm{p}}_2}} \!=\! \textstyle\sum_{k \in \tilde{\mathcal U}}{|{{\bf{h}}_{{\rm{b}},u}^H{{\bf{f}}_{{\rm{p}},k}}} |^2} \!+\! {\sigma _u^2}$. On the other hand, ${{\bf{E}}_{{\rm{c}}_1}} \!=\!{{\tilde{\bf{H}}}_{\rm{c}}^H}{{\bf{h}}_{{\rm{r}},u}}{{\bf{h}}_{{\rm{r}},u}^H} {{\tilde{\bf{H}}}_{\rm{c}}}$, ${{\bf{E}}_{{\rm{c}}_2}} \!=\!{\bf{f}}_{\rm{c}}^H{{\bf{h}}_{{\rm{b}},u}}{{\bf{h}}_{{\rm{r}},u}^H} {{\tilde{\bf{H}}}_{\rm{c}}}$, ${{\bf{E}}_{{\rm{c}}_3}} \!=\! {{\bf{E}}_{{\rm{p}}_1}} \!+\! {{\bf{E}}_{{\rm{p}}_3}}$, ${{\bf{E}}_{{\rm{c}}_4}} \!=\! {{\bf{E}}_{{\rm{p}}_4}}$, ${{\bf{E}}_{{\rm{c}}_5}} \!=\! {{\bf{E}}_{{\rm{p}}_2}} \!+\! {{\bf{E}}_{{\rm{p}}_5}}$, ${\tau _{{\rm{c}}_1}} \!=\! {|{{\bf{h}}_{{\rm{b}},u}^H{{\bf{f}}_{\rm{c}}}} |^2}$, ${\tau_{{\rm{c}}_2}} \!=\! {\tau_{{\rm{p}}_1}} \!+\! {\tau_{{\rm{p}}_2}}$. With the obtained rate splitting vector $\bf{c}$ in problem $\mathrm{P}_{2}$, we denote $\delta_{1,u} \!=\! {2^{R_u^{\rm{min} } \!-\! {c_u}}} \!-\! 1$ and $\delta_2 \!=\!2^{\textstyle\sum_{u \in {\cal U}} \!{{c_u}}} \!-\! 1$. Then, \eqref{eq:problem0B} and \eqref{eq:problem0C} are rewritten as ${\gamma _{{\rm{p}},u}} \! \ge \! \delta_{1,u}$ and ${\gamma _{{\rm{c}},u}} \! \ge \! \delta_2$, respectively. 

With these transformations, problem $\mathrm{P}_{3}$ is cast into a QCQP. By defining ${\bar{\bm{\varphi }}} \!= \!{[{{\bm{\varphi }}^T},1]^T}$, and ${\bar{\bm{\Phi }}} \!=  \! {\bar{\bm{\varphi }}}{{\bar{\bm{\varphi }}}^H}$, we convert it into SDR form and solve it by the SROCR algorithm, i.e.,
\begin{subequations}
\begin{align}
    \mathop {\max }\limits_{ {\bar{\bf{\Phi }}} \succeq 0, \{\! \varGamma \!\}_{{{\rm{P}}_{3}}}^t} 
    &\mathop \{ \varGamma \}_{{{\rm{P}}_{3}}}^t       \label{eq:problem3-1A} \\
    \mathrm{s.t.} \ & {\rm {tr}} ( {{{\bf{Q}}_1} {\bar {\bf{\Phi }}}} ) \!+\! {\eta _1}  \le  \{ \varGamma  \}_{{{\rm{P}}_3}}^t{\tau _q}, \forall q \! \in \!{\cal Q}, \label{eq:problem3-1B}  \\    
    & {\rm {tr}} ( {{{\bf{Q}}_2} {\bar {\bf{\Phi }}}} ) \!+\! {\eta _2} \le P_{{\rm{RIS}}}^{{\rm{max}}}, \label{eq:problem3-2C}  \\
    & {\rm {tr}} (\! {{{\bf{Q}}_3} {\bar {\bf{\Phi }}}} \!) \!+\! {\tau _{{\rm{p}}_1}} \! \ge \! {\delta_{1,u}} ( {\rm {tr}} (\! {{{\bf{Q}}_4} {\bar {\bf{\Phi }}}} \!) \!+\! {\tau _{{\rm{p}}_2}} ), \! \forall u \! \in \!{\cal U}, \label{eq:problem3-1D}  \\ 
    & {\rm {tr}} (\! {{{\bf{Q}}_5} {\bar {\bf{\Phi }}}} \!) \!+\! {\tau _{{\rm{c}}_1}}\! \ge \!{\delta_2} ( {\rm {tr}} ( {{{\bf{Q}}_6} {\bar {\bf{\Phi }}}} ) \!+\! {\tau _{{\rm{c}}_2}} ),\! \forall u \! \in \!{\cal U}, \label{eq:problem3-1E}  \\
    &{\bar {\bf{\Phi }}}_{l,l} \le {a_{\max }^2},\forall l \in \mathcal{L},  \label{eq:problem3-1F} \\
    & {\bf{u}}_{\max }^H( {{\bar {\bf{\Phi }}}^{( {\!t \!-\! 1\!} )}} ){\bar {\bf{\Phi }}}{{\bf{u}}_{\max }}( {{\bar {\bf{\Phi }}}^{( {\!t \!-\! 1\!})}} ) \ge \varpi^{( {\!t \!-\! 1\!} )}{\rm{tr}}( {\bar {\bf{\Phi }}} ),
\end{align}
\end{subequations}
where ${{\bf{Q}}_1} \!\!=\!\! \left[\! {\begin{array}{*{20}{c}}
{{\bar{\bf{Q}}}_1}&{\bf{0}}\\
{{{\bf{0}}^T}}&0 \end{array}} \!\right]$, ${{\bf{Q}}_2} \!\!=\!\! \left[\! {\begin{array}{*{20}{c}}
{{\bar{\bf{Q}}}_2}&{\bf{0}}\\
{{{\bf{0}}^T}}&0 \end{array}} \!\right]$, ${{\bf{Q}}_3} \!\!=\!\! \left[\! {\begin{array}{*{20}{c}}
{{{\bf{E}}_{{\rm{p}}_1}}}&{{\bf{E}}_{{\rm{p}}_2}^H}\\
{{{\bf{E}}_{{\rm{p}}_2}}}&0 \end{array}} \!\right]$, ${{\bf{Q}}_4} \!\!=\!\! \left[\! {\begin{array}{*{20}{c}}
{{{\bf{E}}_{{\rm{p}}_3}} \!\!+\!\! \sigma _{\rm{z}}^2{{\bf{E}}_{{\rm{p}}_4}}}&{{\bf{E}}_{{\rm{p}}_5}^H}\\
{{{\bf{E}}_{{\rm{p}}_5}}}&0 \end{array}} \!\right]$, ${{\bf{Q}}_5} \!\!=\!\! \left[\! {\begin{array}{*{20}{c}}
{{{\bf{E}}_{{\rm{c}}_1}}}&{{\bf{E}}_{{\rm{c}}_2}^H}\\
{{{\bf{E}}_{{\rm{c}}_2}}}&0 \end{array}} \!\right]$, ${{\bf{Q}}_6} \!\!=\!\! \left[\! {\begin{array}{*{20}{c}}
{{{\bf{E}}_{{\rm{c}}_3}} \!\!+\!\! \sigma _{\rm{z}}^2{{\bf{E}}_{{\rm{c}}_4}}}&{{\bf{E}}_{{\rm{c}}_5}^H}\\
{{{\bf{E}}_{{\rm{c}}_5}}}&0 \end{array}} \!\right]$, and ${{\bar {\bf{\Phi }}}^{( {\!t \!-\! 1\!} )}}$ is the optimal solution obtained in the $(t\!-\!1)$-th iteration.

\subsection{Convergence and Complexity Analysis}

\begin{algorithm}[t]
    \renewcommand{\algorithmicrequire}{\textbf{Initialize:}}
	\renewcommand{\algorithmicensure}{\textbf{Output:}}
	\caption{The proposed BCD algorithm} \label{alg1}
    \footnotesize
    \begin{algorithmic}[1] 
        \REQUIRE ${{\bf{F}}^{(0)}}$, ${\bf{\Phi}}^{(0)}$, $\epsilon = 10^{-3}$, $t=0$   
        
        \REPEAT
            \STATE $t = t+1$.            
            \STATE Update ${\bf{w}}_q^{(t)}$ by \eqref{eq:problem1A}, and calculate $\{ \varGamma^{\rm{opt}}  \}_{{{\rm{P}}_1}}^{{(t)}}$. 
            \STATE Update ${\bf{F}}^{(t)}$ by solving \eqref{eq:problem2-2A}, and calculate $\{ \varGamma^{\rm{opt}}  \}_{{{\rm{P}}_2}}^{{(t)}}$. 
            \STATE Update ${\bf{\Phi}}^{(t)}$ by solving \eqref{eq:problem3-1A}, and calculate $\{ \varGamma^{\rm{opt}}  \}_{{{\rm{P}}_3}}^{{(t)}}$. 
        \UNTIL $|\{ \varGamma^{\rm{opt}}  \}_{{{\rm{P}}_3}}^{{(t)}} - \{ \varGamma^{\rm{opt}}  \}_{{{\rm{P}}_3}}^{{(t-1)}}| / \{ \varGamma^{\rm{opt}}  \}_{{{\rm{P}}_3}}^{{(t-1)}} \le \epsilon$

        \ENSURE ${{\bf{F}}^{(t)}}$, $\mathbf{\Phi}^{(t)}$
    \end{algorithmic}
\end{algorithm}

Algorithm~\ref{alg1} employs a BCD framework to update $\mathbf{w}_q$, $\mathbf{F}$, $\mathbf{c}$, and $\mathbf{\Phi}$. In the $\mathbf{F}$- and $\mathbf{\Phi}$-blocks, the SCA surrogates in \eqref{eq:Gamma1} and \eqref{eq:Gamma2} are constructed from the latest feasible $\varGamma$, so the previous point is feasible for the surrogates and the achieved SINR is non-decreasing with SROCR tightening, i.e., $\{ \varGamma \}_{{{\rm{P}}_1}}^{{(t)}} \!\! \le \!\! \{ \varGamma  \}_{{{\rm{P}}_2}}^{{(t)}} \!\! \le \!\! \{\varGamma  \}_{{{\rm{P}}_3}}^{{(t)}} \!\! \le \!\! \{\varGamma  \}_{{{\rm{P}}_1}}^{{(t+1)}}$. SROCR enforces approximately rank-one solutions in both problem $\mathrm{P}_2$ and $\mathrm{P}_3$ without violating these surrogates. Hence, ${\varGamma^{(t)}}$ is non-decreasing and bounded, guaranteeing Algorithm~\ref{alg1} converges to a stationary point of the relaxed problem. Initialization uses a power-feasible ARIS matrix, maximum ratio transmission (MRT)-based BS beamformers, and a QoS-feasible $\mathbf{c}^{(0)}$.

The dominant complexity of Algorithm~\ref{alg1} comes from solving the SROCR-based 
subproblems in \eqref{eq:problem2-2A} and \eqref{eq:problem3-1A}. Using the worst-case complexity of interior-point SDP solvers \cite{9139273}, the overall complexity scales as $\mathcal{O}(I_3(\mathcal{C}_1+\mathcal{C}_2))$, where $\mathcal{C}_1 = \mathcal{O} (I_1 ({\rm{max}}\{M,7U+Q+3 \}^4\sqrt{M}\log(1/\varepsilon_1)))$, $\mathcal{C}_2 = \mathcal{O} (I_2 ({\rm{max}}\{L+1,2U+Q+L+2 \}^4\sqrt{L+1}\log(1/\varepsilon_2)))$. Here $I_3$ is the outer BCD iteration number, $I_1$ and $I_2$ are the inner SROCR iterations, and $\varepsilon_1$, $\varepsilon_2$ denote the solution accuracies. Compared with ARIS-SDMA, the extra cost mainly stems from the extra rate-splitting constraints in the RSMA design.

\begin{table}[t]
	\caption{Simulation parameters} \label{tab:SimPara}
	\centering
    \small
    \begin{tabular}{l|l||l|l||l|l}
		\hline 
		\textbf{Para.} & \textbf{Value} & \textbf{Para.} & \textbf{Value} & \textbf{Para.} & \textbf{Value} \\
		\hline    
        $M$  & 16 & $\alpha_{\rm{b,r}}$  & 2.2  & ${\phi _q}$  & $\{0^\circ,45^\circ\}$  \\
        $L$  & 32 & $\alpha_{{\rm{b}},u}$  & 3.5 & $P_{{\rm{BS}}}^{{\rm{max}}}$  &  40 dBm
        \\
        $U$  & 4  & $\alpha_{{\rm{r}},u}$  & 2.3 & $P_{{\rm{RIS}}}^{{\rm{max}}}$  &  20 dBm  \\
        $Q$  & 2 & $\alpha_{{\rm{r}},q}$  & 2.2  & $\sigma _{u}^2, \sigma _{\mathrm{z}}^2,\sigma _{\mathrm{r}}^2$  & -80 dBm \\       
		\hline
	\end{tabular} 
\end{table}

\section{Numerical Results}

This section evaluates the proposed ARIS-RSMA-assisted ISAC scheme and the joint design algorithm through numerical simulation. The channel model with large- and small-scale fading is similar to \cite{10054402}, where $\alpha_{\rm{b,r}}$, $\alpha_{{\rm{b}},u}$, $\alpha_{{\rm{r}},u}$ and $\alpha_{{\rm{r}},q}$ are the path-loss exponents for the BS-RIS, BS-user, RIS-user, and RIS-target channels, respectively. We set $\delta_i^{(0)}=0.1$ for all SROCR updates. Unless otherwise stated, the specific parameter configurations are listed in Table \ref{tab:SimPara}. To verify the performance superiority of ARIS and RSMA in ISAC systems (ARIS-RSMA), we also provide comparisons with benchmarks based on PRIS, NOMA, SDMA, and sensing-only schemes (OnlyS).

\begin{figure}[t] 
\centering
  \begin{minipage}{0.49\linewidth}
    \centering
    \includegraphics[width=\linewidth]{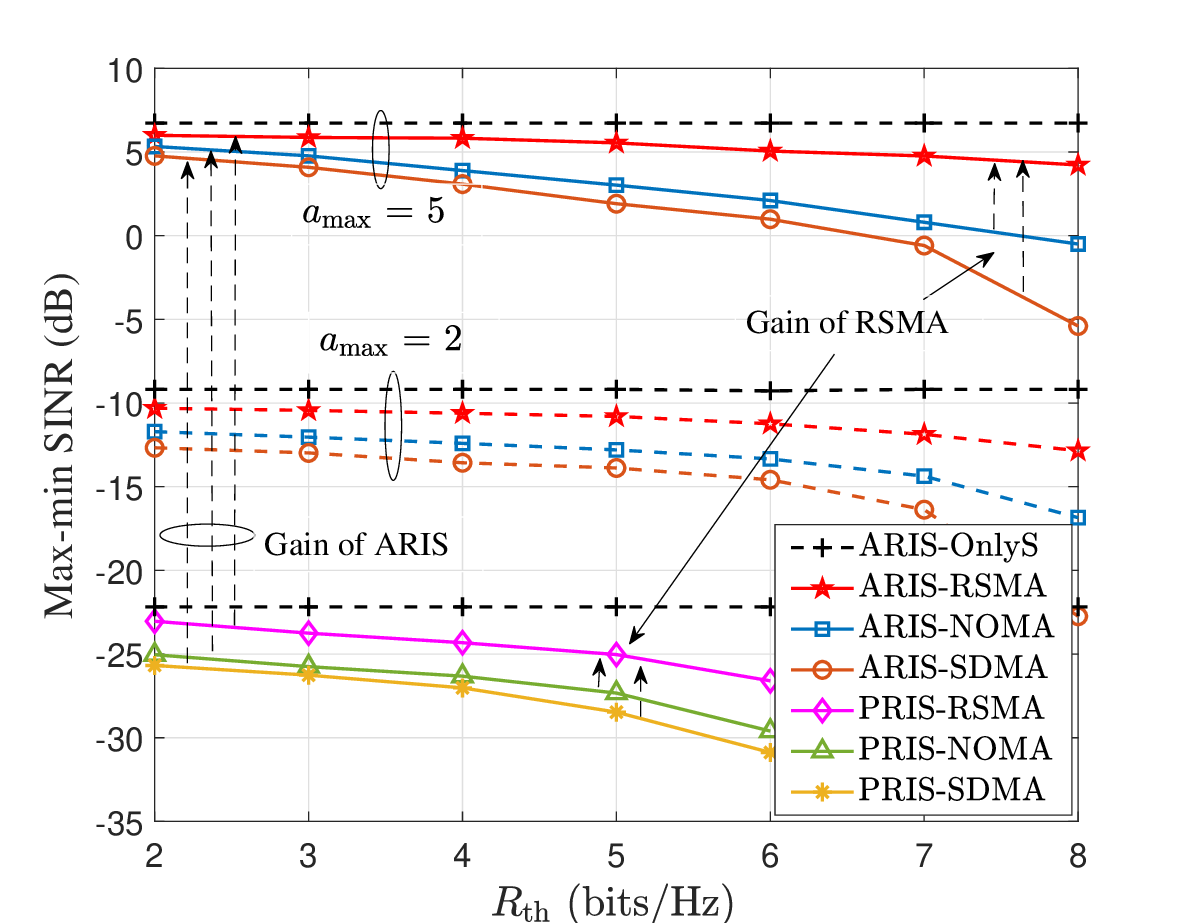} 
    \captionsetup{justification=raggedright,font={small}}
    \caption{Max-min SINR vs. $R_{\rm{th}}$.}
    \label{fig:MaxSINR_Rth}
  \end{minipage}  
  \hfill
  \begin{minipage}{0.49\linewidth}
    \centering
    \includegraphics[width=\textwidth]{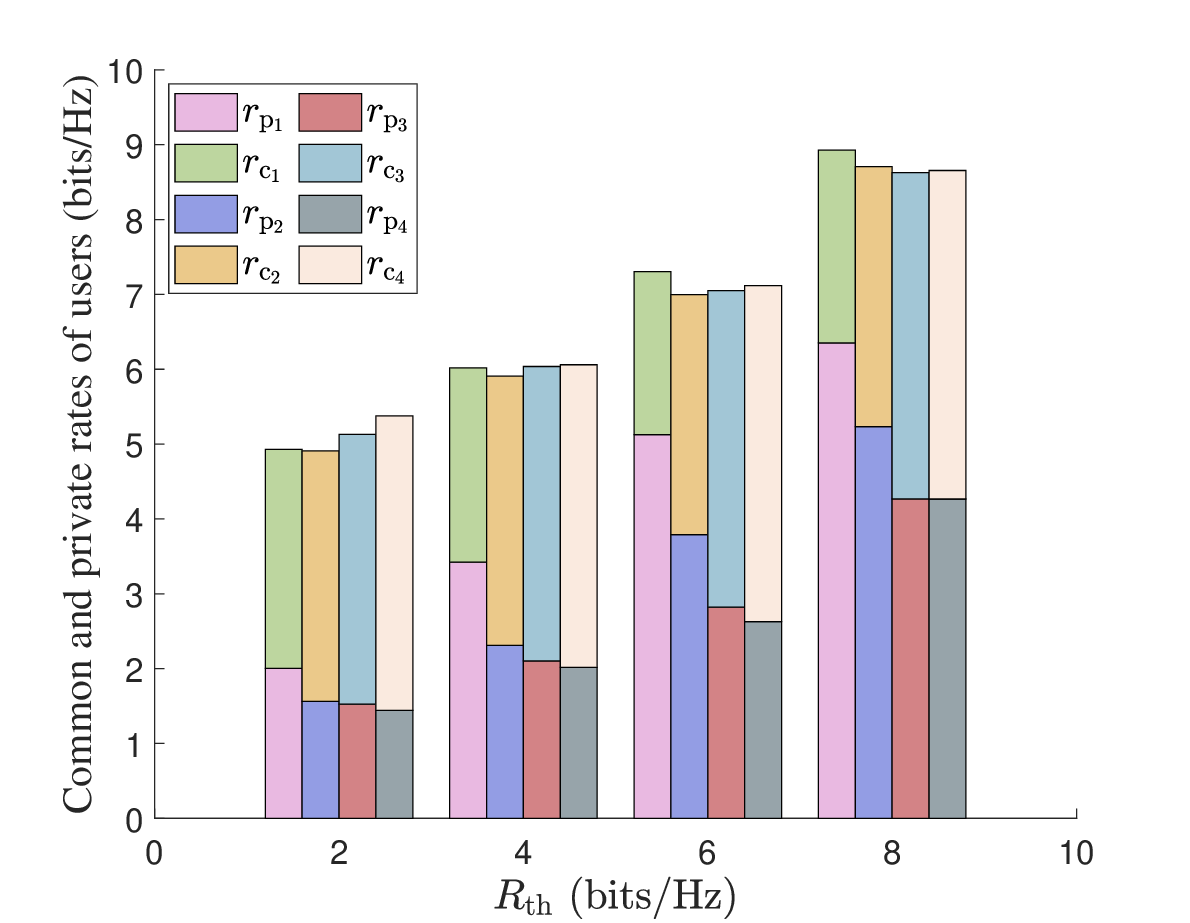}
    \captionsetup{justification=raggedright,font={small}}
    \caption{Rate allocation vs. $R_{\rm{th}}$.}
    \label{fig:RateAllocation}
  \end{minipage}
\end{figure}

\begin{figure}[t] 
\centering
  \begin{minipage}{0.49\linewidth}
    \centering
    \includegraphics[width=\linewidth]{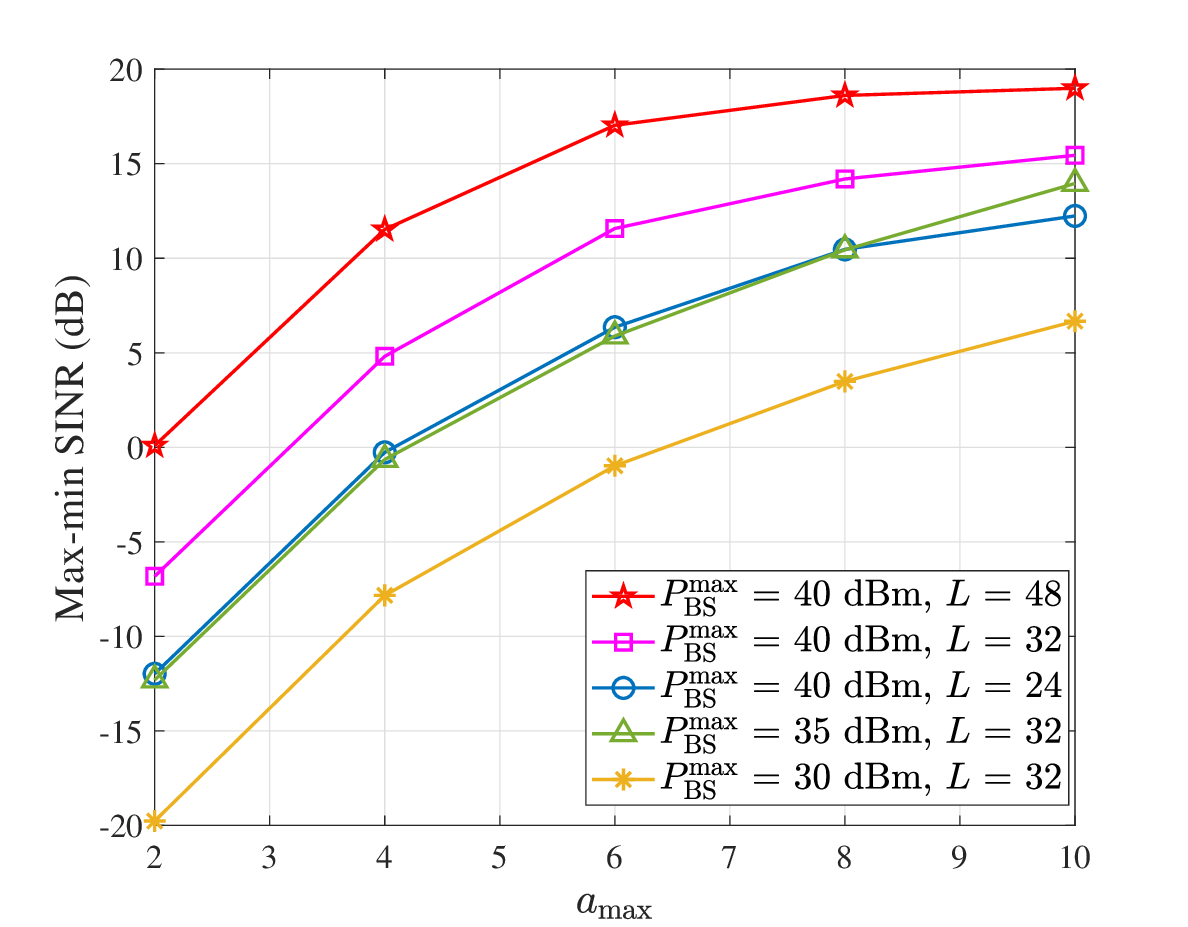} 
    \captionsetup{justification=raggedright,font={small}}
    \caption{Max-min SINR vs. $a_{\rm{max}}$.}
    \label{fig:MaxSINR_amax}
  \end{minipage}  
  \hfill
  \begin{minipage}{0.49\linewidth}
    \centering
    \includegraphics[width=\textwidth]{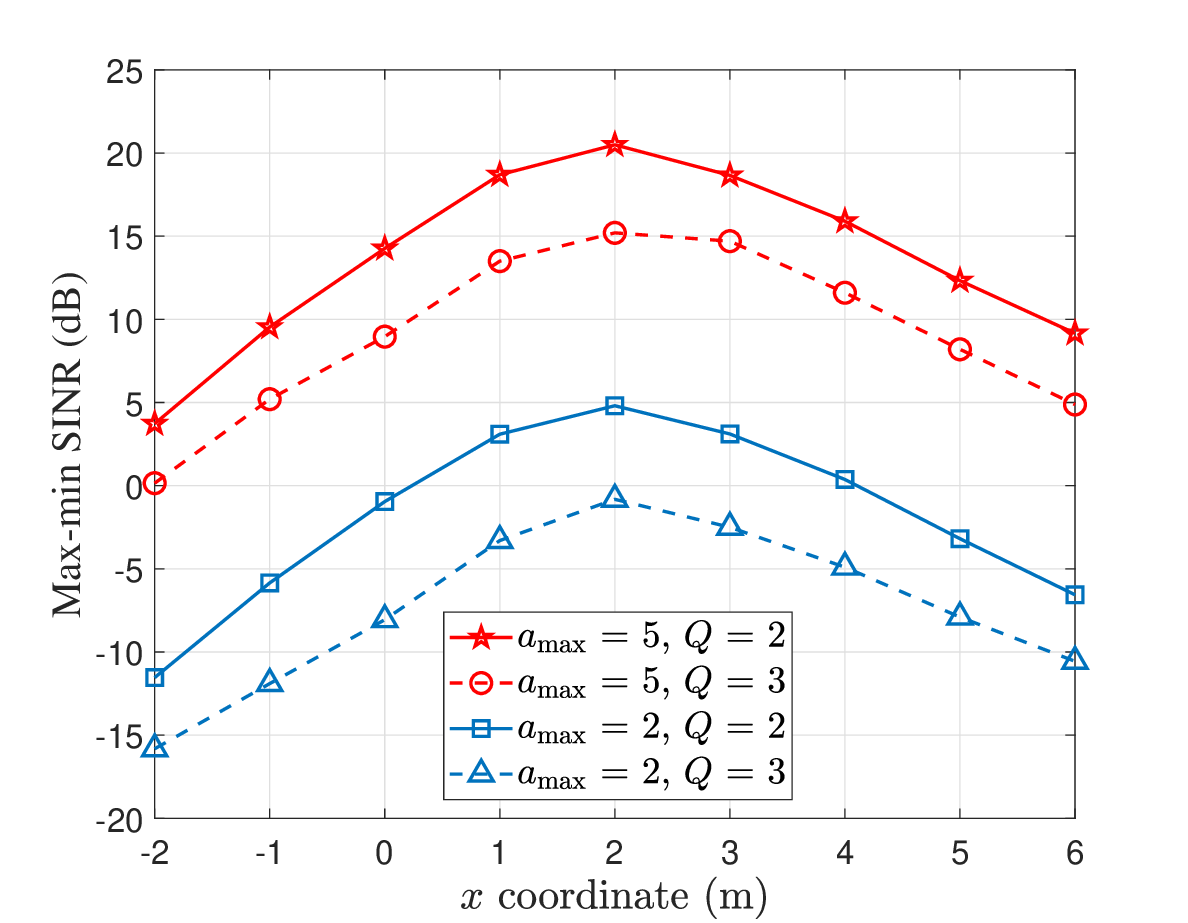}
    \captionsetup{justification=raggedright,font={small}}
    \caption{Max-min SINR vs. ARIS location.}
    \label{fig:MaxSINR_dis}
  \end{minipage}
\end{figure}

Fig. \ref{fig:MaxSINR_Rth} shows the max-min multi-target echo SINR versus $R_{\rm{th}}$ for different schemes. As $R_{\rm{th}}$ increases, all communication enabled schemes suffer a sensing loss, while ARIS always outperforms PRIS. A larger $a_{\mathrm{max}}$ further strengthens the coverage-limited echo link. For a fixed RIS architecture, RSMA yields higher worst-target echo SINR than NOMA and SDMA, e.g., at $R_{\rm{th}}=5$ bits/Hz, 5.54, 3.02, 1.91 dB for ARIS-RSMA, ARIS-NOMA, ARIS-SDMA, respectively. This is because SDMA’s private-only transmission consumes more spatial resources to meet QoS, NOMA’s fixed SIC constraints limit rebalancing under ARIS power and noise coupling, whereas RSMA’s common-private structure best exploits the ARIS-enhanced spatial resources for weakest-target protection.

Fig. \ref{fig:RateAllocation} depicts the common and private rate allocation versus $R_{\rm th}$ for ARIS-RSMA. When $R_{\rm th}$ is small, most user rates are carried by the common stream and only modest private rates are needed, leaving more power and spatial degrees of freedom for sensing. As $R_{\rm th}$ grows, the common rate becomes limited by the weakest user, so the optimizer shifts power to private streams, causing the private-rate share to increase while the common rate saturates and the max-min echo SINR decreases. 

Fig. \ref{fig:MaxSINR_amax} presents the max-min echo SINR versus $a_{\max}$. Given $P_{\mathrm{BS}}^{\max}$ and $L$, increasing $a_{\max}$ improves the echo SINR but with diminishing returns, as the ARIS power constraint and amplified noise gradually limit the effective gain. Higher $P_{\mathrm{BS}}^{\max}$ or larger $L$ shift the curves upward, indicating that a moderate $a_{\max}$, e.g., $5$, is sufficient to harvest most of the ARIS gain without incurring excessive ARIS power consumption.

Fig. \ref{fig:MaxSINR_dis} plots the max-min echo SINR versus the ARIS horizontal coordinate $x$ for different target numbers and amplification factors. All curves exhibit a clear peak around $x \approx 2$ m, indicating that the ARIS should be deployed near, but not necessarily closest to, the blocked sensing region to balance the BS-ARIS-target round-trip gain and the ARIS-assisted user links under QoS constraints. Increasing $a_{\max}$ from $2$ to $5$ and reducing the number of targets from $Q=3$ to $Q=2$ consistently shift the curves upward, which confirms that stronger amplification and lower target density are conducive to safeguarding the worst-target performance.

\section{Conclusion}

This letter presents an ARIS-RSMA-enhanced ISAC framework to address the fairness challenge in multi-target sensing under obstructed LoS conditions. We formulate a joint optimization problem for beamforming and rate splitting, solved via an efficient BCD algorithm, to guarantee worst-case sensing performance across multiple targets. Simulation results demonstrate that the proposed scheme maintains user QoS while significantly improving sensing robustness and fairness, offering an effective solution for ISAC in complex propagation environments.

\appendices

\ifCLASSOPTIONcaptionsoff
 \newpage
\fi

\bibliographystyle{IEEEtran}
\bibliography{Reference}

\end{document}